\documentclass[twocolumn,aps,prb,amsmath,amssymb,floatfix,showpacs]{revtex4}
\usepackage{graphicx}% Include figure files
\usepackage{hyperref}

\begin{document}

\title{Inhomogeneous Phases in a
  Double-Exchange Magnet with Long Range Coulomb Interactions}
\author{Gideon Wachtel}
\affiliation{Racah Institute of Physics,
  the Hebrew University, Jerusalem 91904, Israel}
\author{Denis I. Golosov}
\affiliation{Department of Physics and the Resnick Institute, Bar-Ilan
University, Ramat-Gan 52900, Israel}

\author{Dror Orgad}
\affiliation{Racah Institute of Physics,
  the Hebrew University, Jerusalem 91904,  Israel}
\date{\today}

\begin{abstract}
  We consider a model with competing double-exchange
  (ferromagnetic) and super-exchange (anti-ferromagnetic)
  interactions in the regime where phase separation takes place.
  The presence of a long range Coulomb interaction frustrates a
  macroscopic phase separation, and favors microscopically
  inhomogeneous configurations. We use the variational
  Hartree-Fock approach, in conjunction with Monte-Carlo simulations
  to study the geometry of such configurations in a two-dimensional
  system. We find that an array of diamond shaped ferromagnetic
  droplets is the preferred configuration at low electronic
  densities, while alternating ferromagnetic and
  anti-ferromagnetic diagonal stripes emerge at higher
  densities. These findings are expected to be relevant for thin
  films of colossal magneto-resistive manganates.
\end{abstract}

\maketitle

\section{Introduction}

\label{sec:intro}

In recent years, doped manganese oxides remained at the forefront of
theoretical and experimental research\cite{Tokurabook}. The main
source of interest in theses systems is the phenomenon of colossal
magnetoresistance (CMR), which they exhibit, and that is likely to
have important technological applications. In the meantime, the
underlying basic physics remains elusive, and probably involves the
strongly-correlated nature of the doped magnetic oxides. The CMR in
doped manganates is observed for intermediate hole-doping levels,
typically $0.2 \alt x \alt 0.5$, in the temperature region around
the transition between low-temperature metallic ferromagnetic (FM)
and high-temperature insulating paramagnetic phases. In addition to
double-exchange ferromagnetism\cite{deGennes}, the CMR compounds
also possess pronounced antiferromagnetic (AFM) tendencies, as
evident from the AFM spin ordering with N\'{e}el temperatures of
about 100-200K, observed\cite{Schiffer,neel} at the doping
end-points ($x=0$ and $x=1$). This antiferromagnetism is of a
superexchange origin\cite{virtual}.
% between the core electrons, and for intermediate values of $x$ also
% includes a contribution from virtual transitions between the
% spin-split components of the conduction band\cite{prb05} (owing to a
% finite value of the Hund's exchange coupling, $J_H$, between the
% core and conduction electrons).

The manganates physics involves several degrees of freedom of
substantially different nature, including localized core spins
$\mathbf{S}_i$ of Mn ions, fermionic degrees of freedom associated
with conduction $e_g$-electrons, lattice distortions, {\it etc}. In
such systems, the presence of competing interactions (such as FM and
AFM) often gives rise to phase
separation\cite{Nagaevbook,Dagottobook,psgeneral,Mishra}, whereby areas of
different phases are stabilized in a structurally and
stoichiometrically homogeneous sample. In the case of the
manganates, it has even been suggested\cite{Dagottobook,Uehara} that
phase separation into insulating paramagnetic and metallic FM phases
may explain the resistivity peak observed near the Curie
temperature. In the present paper, we focus on the low-temperature
regime where the presence of phase separation in the appropriate
manganate systems has been directly verified, {\it e.g.,} by means
of scanning tunneling microscopy (STM) on thin films\cite{Biswas}.
Transport measurements reveal metastability and history dependence
near the percolation threshold ($x\approx 0.2$), confirming phase
separation in both film\cite{phasepfilms,Vlakhov} and
crystalline\cite{bgu,Tokunaga} samples.

Using simple microscopic models\cite{deGennes,Nagaev72} it can
readily be shown\cite{Nagaev72} that the hole concentration $x$
indeed controls the balance between the FM and AFM tendencies of the
system. Once $x$ is tuned away from the optimal CMR doping region,
the homogeneous FM metallic state no longer corresponds to the
energy minimum. Instead, energy can be gained by changing the
magnetic ordering, carrier density, bandstructure, and/or orbital
state in {\it part} of the system, making the sample
inhomogeneous\cite{Nagaevbook,Dagottobook,Nagaev72,psgeneral,Mishra,jap02}.
The surface tension between different phases\cite{Nagaev72,prb06}
then competes against the long-range interactions present in the
system in the form of electrostatic
forces\cite{Nagaevbook,Nagaev72,frustrated,prb03} or long-range crystal strain
fields\cite{Ahn,Khomskii}. These require that the system remains
homogeneous at least {\it on average} on the appropriate length
scale (such as the Debye -- H\"{u}ckel screening length), resulting
in a periodic arrangement of nano- or mesoscopic regions of
different phases\cite{Nagaevbook,Dagottobook}. The geometry of the
ensuing inhomogeneous (phase separated) state is at the focus of our
present study.

Early studies of phase separation in double-exchange --
superexchange systems\cite{Nagaevbook,Nagaev72,Nagaev00} implicitly
assumed that the effects of the discrete lattice are unimportant,
and consequently treated the problem within the continuum based,
long-wavelength, approach. Within this framework, the optimal phase
separation geometry at small values of the FM volume (or area) fraction
$m$ (also the average magnetization per site) is obviously that of
spherical (in two dimensions, circular) FM droplets located at the
sites of a packed hexagonal (triangular) super-lattice. To the best
of our knowledge, only the three-dimensional case was treated in
detail, with the implication that in two dimensions the situation is
similar.  When the system parameters are varied in such a way that
$m$ increases beyond $1/2$, the geometry changes to that of
spherical AFM droplets in an otherwise FM matrix. The change
generally occurs via a direct ``geometrical phase
transition''\cite{Nagaev00} without any intervening regime
characterized by both phases forming infinite connected shapes (such
as filaments and planar slabs in three dimensions or stripes in two
dimensions)\cite{Nagaevbook,Nagaev72,3dslabs}.

This latter conclusion is important, since such slab or stripe
arrangements, if realized, would have been characterized by peculiar
and potentially useful properties such as history-dependent
anisotropy of the ground state resistivity.  However, the continuum
treatment, which is at the basis of this result, is not valid beyond
the region of very small values of $x \ll 1$. Indeed, recent studies
suggest\cite{jap02,prb06} that the boundary between the two phases
is abrupt on the lattice-spacing scale ({\it i.e.,} of the type
commonly associated with Ising spin systems). Such a boundary cannot
be adequately described in the continuum limit, and its surface tension
depends on its orientation with respect to the crystalline axes\cite{prb06}.
% even if the latter is a cubic or square one.
This directional dependence of boundary energies should in turn
affect the droplet shape (generally favoring diamond-shaped droplets
in two dimensions)\cite{prb06}, the arrangement of droplets in
space, and ultimately the way the geometry of phase separation
evolves with varying $m$. This is apparently a generic property
of electronic phase separation, found also within the frameworks of
Falikov--Kimball\cite{Nachtergaele} and $t-J$ (Ref.
\onlinecite{Richard}) models.

In the present paper, we revisit the problem within the framework of
a single-orbital double-exchange -- superexchange Hamiltonian (with
infinite Hund's coupling $J_H$), augmented by a long-range Coulomb
interaction term. Using a variational Hartree-Fock approach, we
compute the energies of various two-dimensional droplet and stripe
phases corresponding to a FM area fraction $m \lesssim 1/2$,
and determine the optimal configuration. Our most important finding
is that while a droplet lattice exists at low doping levels, a
striped arrangement has a lower energy and is therefore stabilized
over a broad region of the phase diagram. As anticipated from our
previous results concerning the orientational dependence of the
boundary energy\cite{prb06}, we find that diamond droplets and
diagonal stripes are the preferred geometries for the FM regions of
the inhomogeneous states. These conclusions gain further support
from unrestricted Hartree-Fock calculations which we have carried
out using Monte-Carlo simulated annealing on moderate size clusters.
The simulations also demonstrate the existence of inhomogeneous
states comprised of AFM droplets (or stripes) embedded in a FM
background ($m \gtrsim 1/2$), at higher doping levels. While
our results pertain to the two-dimensional case, it is likely that
qualitatively our conclusions would also apply to three-dimensional
systems. Specifically, we suggest that a phase separated state with
filament or slab geometry (rather than a lattice of droplets) is
realized for a certain range of parameters in three dimensions.

%For a two-dimensional system, we compute, the energies of
%triangular and square-lattice arrangements of ferromagnetic droplets
%as well as the energy of the striped arrangements in order to find the
%optimal configuration.  We consider the case where the individual
%droplets are either diamond- or square-shaped -- an
%expectation\cite{prb06} which is further corroborated by our present
%analysis.  Most importantly, we find that in the broad region of
%parameter values corresponding to ferromagnetic area fraction $\Delta
%\alt 1/2$, the {\it striped arrangement has a lower energy and is
%  therefore stabilised in place of a droplet lattice}. Our calculation
%of droplet-phase energies is based on a Hartree--Fock treatment of the
%many-body problem for holes and is valid for the case of disconnected
%metallic droplets in the insulating matrix (relevant for $\Delta
%<\Delta_c$ with a certain $\Delta_c \approx 1/2$). The conclusions,
%however, are confirmed by direct numerical simulations spanning all
%values of $\Delta$.

In addition, we find that the typical droplet size and stripe width
do not exceed several lattice constants. This means that the motion
of the charge carriers is strongly quantized, rendering droplets
midway between metallic bulk and magnetic polarons\cite{Nagaev67}
and giving rise to singularities in the stripe energy associated
with the quantisation of the transverse kinetic energy. This
important property was not included in the earlier
work\cite{Nagaevbook,Nagaev72,Nagaev00}, which assumed sufficiently
large length scales for such quantum effects to be negligible. Our
approach, on the other hands, allows one to explore the crossover
between the regime of singly-occupied magnetic polarons, which
appear for strong Coulomb and AFM couplings, and the more
conventional phase separation behavior where each metallic droplet
is populated by many charge carriers.

The paper is organized as follows: in Sec. \ref{sec:model}, we
introduce the model and briefly review the physics underlying phase
separation and magnetic polaron formation in the absence of a
long-range force. A brief description of the calculational methods
which were implemented in order to include the effects of the
long-range Coulomb repulsion appears in Sec. \ref{sec:methods},
while the mass of details is relegated to the appendices. Sec.
\ref{sec:results} contains a detailed description of our
Hartree--Fock and Monte-Carlo results. We conclude with a brief
discussion of the results in the context of current experimental and
theoretical work (Sec. \ref{sec:conclu}). While an arrangement of
conducting and insulating stripes in doped manganate films has not
yet been observed, we suggest that present experimental knowledge
should allow for a meaningful and successful research effort in this
direction.

\section{The Model and its Properties in the Non-Interacting Limit}

\label{sec:model}

The starting point for the following calculation is the
two-dimensional double-exchange Hamiltonian, generalized to include
the superexchange coupling and the long-range Coulomb interaction,
\begin{eqnarray}
  \!\!\!\!\!\!\mathcal{H}&=& -\frac{t}{2}\sum_{\left\langle i,j\right\rangle \alpha}
  \left(c_{i\alpha}^{\dagger} c_{j\alpha} + \mbox{H.c.}\right)
  +\frac{J}{S^2}\sum_{\left\langle i,j\right\rangle }\mathbf{S}_{i}\cdot
  \mathbf{S}_{j}\nonumber \\
  &-&\frac{J_{H}}{2S}\sum_{i,\alpha,\beta}\mathbf{S}_{i}\cdot
  \left(c_{i\alpha}^{\dagger}\boldsymbol{\sigma}_{\alpha\beta}
    c_{i\beta}\right) \nonumber \\
  &+&U\sum_{i\ne j,\alpha,\beta}\frac{1}{\left|\mathbf{r}_{i}-\mathbf{r}_{j}\right|}
  \left(c_{i\alpha}^{\dagger}c_{i\alpha}-x\right)
  \left(c_{j\beta}^{\dagger}c_{j\beta}-x\right).
  \label{eq:Ham_orig}
\end{eqnarray}
Here $t$ is the nearest-neighbor hopping amplitude and $c_{j\alpha}$
annihilates a conduction electron of spin
$\alpha=\uparrow,\downarrow$ at site $j$ of a square lattice.
$\mathbf{S}_{i}$ denotes the core spin made of three d-shell
electrons ($S=3/2$) localized at site $i$, whose AFM superexchange
interaction with neighboring core spins is given by the second term
in ${\cal H}$. The third term arises from Hund's coupling between
the core spins and the conduction electrons, where the spin operator
for the conduction electrons on site $i$ involves the Pauli matrices
$\boldsymbol{\sigma}$. It is this term, in conjunction with the fact
that the hopping preserves the electronic spin, which gives rise to
the double-exchange mechanism. This favors a FM spin configuration in
order to reduce the conduction electrons' kinetic
energy\cite{deGennes}. The last term includes the Coulomb
interaction among the conduction electrons, whose average density is
$x$, and a neutralizing uniform positive background, created by the
donors.
%sentences added by DG:
Owing to the long-range nature of the Coulomb interaction, the
atomic-scale inhomogeneities of this background in real systems (created
by chemical substitution) are not expected to be important
from the point of view of our main purpose of comparing the 
energies of various inhomogeneous phases. This is because such energies
always involve integration over volume.

%to be inessential, at least from the
%point of view of our main purpose of comparing the energies of various
%inhomogeneous phases.
%This is because we are primarily interested in evaluating
%energies of different phases, {\it i.e.,} integral quantities.
% $\frac{1}{2}\sum_{\alpha\beta}c_{i\alpha}^{\dagger}\boldsymbol
% {\sigma}_{\alpha\beta}c_{i\beta}$, $\boldsymbol{\sigma}$ being the
% Pauli matrices.  The combination of these two terms encompasses the
% double exchange mechanism, which favours a ferromagnetic
% configuration in order to reduce the conduction electrons' kinetic
% energy.  The other two terms are the anti-ferromagnetic super
% exchange coupling between neighboring core spins, and the long range
% Coulomb interaction between conduction electrons. Also included is
% the Coulomb interaction with a neutralizing uniform positive
% background, due to the presence of the donors.

%New paragraph added by DG
In using the simplified model, Eq. (\ref{eq:Ham_orig}), we neglect
some additional physics characteristic of the CMR manganates\cite{Tokurabook}.
This includes the presence of two (rather than one) conduction electron
$e_g$-bands and the electron-lattice coupling. The logics behind
this simplification is summarized, {\it e.g.,} in Ref. \onlinecite{Mishra}:
it is assumed that the mechanism for phase separation (charge ordering)
is the competition between ferro- and antiferromagnetism in the presence
of a long-range Coulomb repulsion [all contained in Eq. (\ref{eq:Ham_orig})].
Once the charge ordering is established, in a real system the orbital
ordering (and the lattice distortions) would follow, leading to a
quantitative renormalization of the parameter values. The model,
Eq. (\ref{eq:Ham_orig}), is however expected to suffice for a qualitative
study of the generic features of the phase diagram while its simplicity
allows to maintain clarity of analysis. Further arguments regarding the
expected model-independence of our conclusions shall be given in Sec.
\ref{sec:conclu}.

The relatively large value $S$ of the Mn core spins means that their
fluctuations are small, particularly in the $T \rightarrow 0$ limit
considered here. In the following we assume $S\gg1$ and treat the
core spins classically. Consequently, the effective Hamiltonian
governing the physics of the conduction electrons is determined by
the configuration of the classical spins $\{\mathbf{S}_i\}$. As far
as the Hund's coupling is concerned the manganates are characterized
by a moderate bare $J_H \lesssim t$. However, they also include a
strong Hubbard on-site repulsion, $U_0 \gg t$, which significantly
renormalizes $J_H$ towards the strong coupling limit\cite{prb05}.
Therefore, while we omit the Hubbard interaction from our
Hamiltonian (\ref{eq:Ham_orig}), we model its effects by taking
$J_{H}\to\infty$.

%Another new paragraph added by DG:
Band theory calculations\cite{Mishra,band} suggest that typical values
of the hopping amplitude $t$ in the CMR manganates lie between 0.3 eV
and 0.5 eV.
The value of
$J$ can be roughly estimated
from the experimentally observed N\'{e}el temperatures in the fully doped
or undoped (with no conduction $e_g$ electrons or with no holes) 
case\cite{Schiffer,neel}, $T_N \sim 100-200K$, corresponding to
$J \sim 5-10 {\rm meV}$. The long-range Coulomb interaction strength, $U$,
for thin films is evaluated as $U=e^2/(a_0 \bar{\epsilon})$. Here, $e$ is the
electron charge and
$a_0 \approx 3.9$ \AA ~is the lattice spacing. The {\it effective} dielectric
constant $\bar{\epsilon}$ is given by the average of dielectric constant
$\epsilon_s$ of the substrate and that of the air,
$\bar{\epsilon}=(\epsilon_s+1)/2$. Among the substances which can be used
as substrates for manganate films, lanthanum aluminate and neodymium
gallate
have\cite{substrate,laongo} $\epsilon_s \approx 23$ and 
$\epsilon_s \approx 20$,
yielding $U \approx0.31$ eV
and $U\approx 0.35$ eV  respectively. Dielectric properties of the third
possible substrate, strontium titanate, are strongly dependent on temperature,
with\cite{substrate,sto}
$\epsilon_s$ changing from 24000 (corresponding to $U\approx 0.31$ meV)
at 4.2K to  $\epsilon_s \approx 277$ ($U\approx 0.027$ eV) at 300K. This
suggests a possibility of experimentally varying the value of $U$ by using
different substrates and/or changing temperature.

%Assuming $S\gg1$, the core spins may be treated classically, and their
%fluctuations freeze out in the $T \rightarrow 0$ case considered
%here. Therefore, the effective Hamiltonian governing physics of the
%conduction electrons is determined by the given configuration of the
%classical spins $\{\mathbf{S}_i\}$.  In the case of manganates, the
%actual system is characterised by a moderate value of $J_H \alt t$ and
%strong Hubbard repulsion on-site, $U_0 \gg t$ [not included in the
%Hamiltonian, Eq. (\ref{eq:Ham_orig})] (see Ref. \onlinecite{prb05} and
%references therein).  Nevertheless, it can be expected that for our
%present purposes it is sufficient to study a simplified model
%(\ref{eq:Ham_orig}) in the limit of large Hund's rule coupling
%$J_{H}\to\infty$ (which approximately corresponds to taking $U_0$ into
%account at the mean-field level \cite{prb05}).

Theoretical investigations of the double-exchange -- superexchange
competition have a history of more than 40 years. It is by now
firmly established\cite{Nagaevbook,Dagottobook,jap02} that this
competition is resolved not via a second-order phase transition from
the FM state to a uniform state with a helical or canted magnetic
ordering, but rather via separation of the sample into regions
characterized by different spin arrangements and conduction electron
bandstructures. We will be interested in the case of phase
separation into FM and AFM regions with abrupt Ising-type
boundaries\cite{jap02,prb06} between them. This means that the
resulting configuration of $\mathbf{S}_{i}$ remains collinear, with
all core spins either parallel or anti-parallel to a selected
direction. Thus, it is possible to denote a spin state simply by
$\mathbf{S}_i/S\equiv S_{i}=\pm1$, and the Hamiltonian of the
conduction electrons becomes a function of $\{ S_{i}\}$.  The large
Hund's exchange coupling then forces the conduction electrons' spins
to polarize in parallel with the core spins,
resulting\cite{Anderson} in the following distribution of hopping
amplitudes for a given spin configuration $\{S_i\}$:
\begin{equation}
  t_{ij}\left(\left\{ S_{i}\right\} \right)=\left\{ \begin{array}{ccl}
      -t &  & i,j\mbox{ nearest neighbors and } S_{i}=S_{j}\\
      0 &  & \mbox{otherwise}\end{array}\right. .
      \label{teff}
\end{equation}
After these simplifications, the Hamiltonian takes the form
\begin{eqnarray}
  \mathcal{H}\left(\left\{ S_{i}\right\} \right) &=&
  \frac{1}{2}\sum_{\left\langle i,j\right\rangle }
  \left(t_{ij}\left(\left\{ S_{i}\right\} \right)c_{i}^{\dagger}c_{j}
    +\mbox{H.c.}\right)
  + J\sum_{\left\langle i,j\right\rangle }S_{i}S_{j} \nonumber \\
  &+&U\sum_{i\neq j}\frac{1}{\left|\mathbf{r}_{i}-\mathbf{r}_{j}\right|}
  \left(c_{i}^{\dagger}c_{i}-x\right)
  \left(c_{j}^{\dagger}c_{j}-x\right).
  \label{eq:Ham_simp}
\end{eqnarray}
%Hence, the energy of a given spin configuration $\left\{S_{i}\right\}$
%is given by the sum of the quantum ground state energy of the
%corresponding (conduction electrons') many body problem and the
%classical AFM term.
The model, Eq.  (\ref{eq:Ham_simp}), on the (bi-partite) square
lattice is invariant under the particle-hole transformation
$c_{i}^{\dagger}\rightarrow (-1)^p c_i$, and $x\rightarrow 1-x$,
where $p$ takes the values 0,1 on the two sublattices. As a result
we note that in the following, $x$ acquire the more general meaning
of a carrier density, i.e., either the electronic density or the
hole density relative to the half-filled state.
%We note that the physically relevant case of moderate {\it hole}
%doping is described by the same Hamiltonian (\ref{eq:Ham_simp}) where
%$c_{i}^{\dagger}$ are (spinless) hole creation operators, and $x<1/2$
%is the average {\it hole} density relative to the half-filled state.
We will now briefly review the ground state properties of the
Hamiltonian (\ref{eq:Ham_simp}) at $U=0$.

When the carrier density $x$ is finite, the ground state of the
system at $J\rightarrow 0$ is uniform and FM. With increasing $J$
beyond a certain critical value $J_c(x)$, this uniform FM state
eventually becomes destabilized, and a non-uniform ground state is
obtained instead. In this {\it phase separated} state only part of
the system is occupied by the FM phase. We will be interested in the
case in which the other part is a simple N\'{e}el antiferromagnet
with zero charge-carrier density. A variational study\cite{jap02}
shows that in two dimensions such a phase separated state may be
realized only for $J<J^* \approx 0.036$. At higher values of $J$ the
magnetic ordering in either the electron-rich or the electron-poor
regions of the sample differs from that of a ferromagnet or a
N\'{e}el antiferromagnet.
%(realized at moderate values of $x$, $x\alt 0.245$ for a
%two-dimensional system at $J=J_c(x)$ [Ref. \onlinecite{jap02}]) when
%ensuing magnetic ordering in the other part of the system is that of a
%simple N\'{e}el antiferromagnet, corresponding to zero carrier
%bandwidth and hence to zero carrier density.

\begin{figure}[h]
  \centering
  \includegraphics[width=\linewidth]{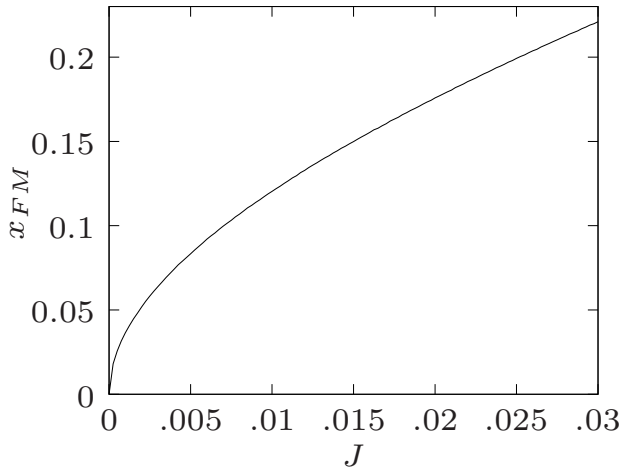}
  \caption{The charge carrier density $x_{FM}$ in the FM region of a macroscopically
  phase separated state.}
  \label{fig:xfmtd}
\end{figure}
Thermodynamic equilibrium between macroscopic FM and AFM regions
means that the thermodynamic potentials in the two phases are equal,
\begin{equation}
  \Omega_{FM}=\Omega_{AFM},
\label{eq:tdstability}
\end{equation}
where
\begin{equation}
  \Omega_{FM} = \int_{-\infty}^\mu(\varepsilon-\mu) g(\varepsilon)d\varepsilon+2J ,
\end{equation}
and
\begin{equation}
  \Omega_{AFM} = -2J .
\label{eq:omegaafm}
\end{equation}
$g(\varepsilon)$ is the density of conduction electron states in the
FM region. Since the latter is large, $g(\varepsilon)$ can be taken
to be the two-dimensional tight-binding density of states, and
boundary corrections may be neglected. By solving Eqs.
(\ref{eq:tdstability}--\ref{eq:omegaafm}) for the Fermi energy,
$\mu$, one readily obtains the carrier density in the FM region,
$x_{FM}$, via
\begin{equation}
  x_{FM}=\int_{-\infty}^\mu g(\varepsilon)d\varepsilon,
\label{eq:tdxfm}
\end{equation}
with the result depicted in Fig. \ref{fig:xfmtd}. The system remains
in a uniform FM state as long as $x_{FM}(J)<x$. The fraction of the
system area (or volume), occupied by the FM phase, is given by
$m=x/x_{FM}$. The critical value, $J_c$, for the onset of phase
separation is then determined by the condition $x_{FM}(J_c)=x$.

%With increasing $J$, the uniform FM state becomes unstable
%with respect to macroscopic phase separation into FM and
%N\'{e}el AFM regions once the value $x_{FM}(J)$ exceeds
%$x$ (Ref. \onlinecite{jap02}).  We note, however, that in addition to
%this macroscopic (thermodynamic) phase separation there also exists a
%different, essentially quantum-mechanical way to resolve the double
%exchange-superexchange competition.

%Beside the global phase separation route, another potential resolution
%to the FM-AFM tension has been considered in the literature, namely, the
%formation of magnetic polarons.
In addition to the macroscopic phase separation as described above,
the double exchange-superexchange competition can also be resolved via
an altogether different scenario (formation of magnetic
polarons). When a single electron (or hole) is lodged into an
antiferromagnetically ordered double-exchange magnet with zero
charge-carriers, ($x=0$), a free (self-trapped) magnetic polaron, or
ferron\cite{Nagaev67,Nagaevbook,polarons,Satpathy01,Pereira,SatpathyCMR,Umeharapol},
is formed around it. It is essentially a microscopic FM region,
containing one charge carrier, in an otherwise AFM system. Since the
propagation of charge is unimpeded in the FM region it acts as a
potential well for the sole carrier, which occupies the lowest bound
state inside the well. The polaron binding energy, $E_{mp}$, (with
respect to the state where the AFM order is unperturbed) can be be
easily estimated\cite{Nagaev67}. We consider the case of a
diamond-shaped polaron , with $L+1$ sites along each side (see Fig.
\ref{fig:geometry}, upper left). For $L \gg 1$ we find
\begin{equation}
E_{mp}(L)=-2t+\frac{t}{2} \left(\frac{\pi}{L+1}\right)^2+8L^2J,
\label{polenergy}
\end{equation}
where the first two terms are the ground-state energy of the charge
carrier and the last one represents the superexchange contribution.
Expression (\ref{polenergy}) should be minimized with respect to $L$,
resulting in
\begin{equation}
E_{mp} =-2t+4 \pi \sqrt{Jt}.
\end{equation}
Here, the coefficient of the second term depends on the geometry of
the FM micro-region ({\it e.g.}, for a round polaron one would have
obtained $12.06$ instead of $4\pi$).

The above expressions are valid in the $J/t \ll 1$ regime (yielding
$L\sim (t/J)^{1/4} \gg 1$), where it is easy to verify an important
statement which is expected to hold for all $J$. Namely, if in the
absence of a Coulomb interaction $U$, a second carrier is added to
the system, it is energetically favorable for the two charge
carriers to occupy the two lowest bound states in a {\it shared} FM
micro-region rather than to form two independent polarons. This
conclusion is verified by calculating the binding energy of the
(diamond-shaped) doubly-occupied polaron
\begin{equation}
E^{(2)}_{mp} =-4t+4\pi\sqrt{7Jt/2},
\end{equation}
which clearly satisfies $E^{(2)}_{mp}<2E_{mp}$. This trend continues
when further charge is added, and at $n \gg 1$ the binding energy
(per carrier) of the $n$-carrier polaron decreases toward the
limiting value $E_{ps}$,
\begin{equation}
\frac{1}{n}E^{(n)}_{mp} \rightarrow E_{ps}\, ,
\end{equation}
which is the energy gain per carrier associated with the macroscopic
phase separated state. The latter can be evaluated as
\begin{equation}
E_{ps}=\frac{E_{FM}-E_{AFM}}{x_{FM}},
\end{equation}
where $E_{FM}(\mu)=\int_{-\infty}^\mu \varepsilon
g(\varepsilon)d\varepsilon+2J$ and $E_{AFM}=-2J$ are the energies
per site of the FM and AFM phases.
% and $N_{FM}=N x/x_{FM}$ is the number of sites in the FM region.
Using Eqs. (\ref{eq:tdstability}) and (\ref{eq:tdxfm}) to evaluate
$\mu$ and $x_{FM}$ we find, in the limit $J \ll t$,
\begin{equation}
E_{ps} = -2t+ 4\sqrt{\pi J t}.
\end{equation}
The inequality, $E_{ps}< E_{mp}$, implies that for any finite
carrier density, at $U=0$, the double-exchange -- superexchange
competition is resolved via macroscopic phase separation.

Notwithstanding the preceding discussion, its conclusion may change
if a realistically strong Coulomb interaction $U$ is included,
favoring a large spatial separation between the charge carriers.
Indeed, as we demonstrate in the following, a polaronic state arises
in the regime of large $U$ and small carrier density. It is the
extreme limit of a broad range of inhomogeneous states which
originate from the frustration of macroscopic phase separation by
long-range forces. The study of this intermediate region of
parameters lies at the focus of the remaining part of the paper.
Since the typical size of the resulting FM regions is rather small
one needs to take into account the effects of quantization of the
charge carrier motion. At the same time, some of the results
obtained for the macroscopic phase-separated system, such as the
directional dependence of the boundary energy\cite{prb06}, still
offer important guidance to the understanding of the inhomogeneous
configurations. Next, we outline the methods used to treat this
intermediate regime which is characterized by a combination of both
traditional phase separation and magnetic polaron (quantized)
physics.

\section{Methods}
\label{sec:methods}

\subsection{Variational Hartree-Fock Approach}

\begin{figure}[ht]
  \centering
  \begin{tabular}{| c | c |}
    \hline
    FM Region Shape &
    Super-Lattice Structure  \\
    \hline
    \includegraphics[scale=1.2]{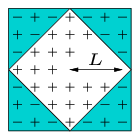}
    \includegraphics[scale=1.2]{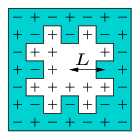}   &
    \includegraphics[scale=1.2]{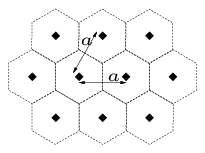}
    \includegraphics[scale=1.2]{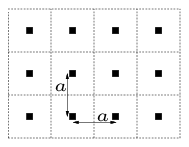} \\
    \hline
    \includegraphics[scale=1.2]{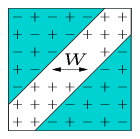}
    \includegraphics[scale=1.2]{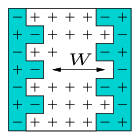}   &
    \includegraphics[scale=1]{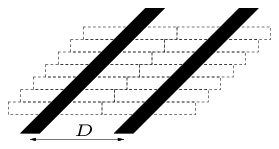}
    \includegraphics[scale=1]{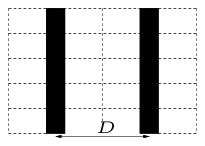} \\
    \hline
  \end{tabular}
  \caption{The geometry of the inhomogeneous
    configurations considered in the variational Hartree-Fock
    calculation. The dashed lines indicate the super-lattice unit
    cell.}
  \label{fig:geometry}
\end{figure}

Given the Hamiltonian (\ref{eq:Ham_simp}), our task is to find the
configuration of core spins in the ground state. However, there is a
vast multitude of possible spin configurations amongst which the
ground state is to be sought, making it impossible to explore all of
them. Nevertheless, previous studies of similar or related
systems\cite{DEnumerical} suggest several families of highly
symmetrical configurations as reasonable ground state candidates.
The two main types of spin configurations studied in this work are
FM droplets in an AFM background, and alternating FM and AFM
stripes, as illustrated in Fig. \ref{fig:geometry}. A uniform FM
phase, in which the double-exchange mechanism completely overcomes
the superexchange, is also considered.

Calculating the energy of the conduction electrons in a given
configuration of core spins is a difficult problem. Here we suffice
with the Hartree-Fock (HF) approximation, which gives an upper-bound
to their ground state energy. Since we are dealing with periodic
spin configurations, the HF equations for the whole system can be
rewritten as an effective eigenvalue problem within a single unit
cell. The superexchange contribution to a configuration's energy is
simply calculated by counting the number of FM and AFM bonds in a
unit cell.

Based on previous analytical results\cite{prb06} and numerical
investigations\cite{DEnumerical} the considered droplets are either
diamond or square shaped, and are chosen to form either a triangular
or square super-lattice (see Fig. \ref{fig:geometry}). Several
droplet phases are possible by combining different droplet shapes
and super-lattice types. In addition, one has variational freedom to
specify $L$, the size of the FM droplets, and $n$, the number of
conduction electrons in each one of the droplets. The distance
between the droplets, $a$, is then uniquely determined by the type
of super-lattice, by $n$, and by the average density of conduction
electrons, $x$.

The energy of each droplet phase is found by minimizing its energy
density,
\begin{equation}
  E_{droplet}=x\,\frac{E_{HF,droplet}}{n}+E_{J,droplet},
\end{equation}
with respect to the variational parameters, $L$, and $n$. Here
$E_{HF,droplet}$ is the HF energy of the conduction electrons inside
a unit cell containing a single droplet, and $E_{J,droplet}$ is the
AFM energy \emph{per site}.
The details of the HF calculation appear in Appendix
\ref{app:hartree}. The main source of complication is the necessity
to take into account the Hartree interaction between electrons
belonging to different droplets in the infinite super-lattice. This
is done by employing Ewald's summation method (see Appendix
\ref{app:ewald}). The AFM coupling energy per lattice site is
\begin{equation}
  E_{J,droplet}=\left\{ \begin{array}{lcl}
      -2J\left(1-\frac{4L^2}{A}\right)
      &  & \!\!\mbox{\normalsize diamond droplets}\\
      -2J\left(1-\frac{5L^2-8L+8}{A}\right)
      &  & \!\!\mbox{\normalsize square droplets}\end{array}\right.
\end{equation}
where $A = n/x$ is the area of a unit cell.

Two types of stripe phases were considered: diagonal and
bond-aligned. Additional variational freedom comes from the need to
specify $W$, the stripe width, and $x_{FM}$, the (average) density
of conduction electrons within the FM stripe. Just as for the
droplet phase, the stripe phase energy is found by minimizing its
energy density
\begin{equation}
  E_{stripe} = x\,\frac{E_{HF,stripe}}{\lambda} + E_{J,stripe}
\end{equation}
with respect to $W$ and $x_{FM}$. Here $\lambda$ and $E_{HF,stripe}$
are, respectively, the conduction electrons' number and energy per
unit cell, and $E_{J,stripe}$ is the AFM energy per site.  A unit
cell in diagonal stripes is only one lattice spacing long in the
direction along the stripe, and two spacings long in bond-aligned
stripes (see Fig. \ref{fig:geometry}); its width equals the stripe
periodicity. Therefore,
\begin{equation}
  \lambda=\left\{ \begin{array}{lcl}
      x_{FM}W &  & \mbox{diagonal stripes}\\
      2x_{FM}W &  & \mbox{bond-aligned stripes}\end{array}\right.\;.
\end{equation}
$\lambda$, together with $x$, uniquely determine the distance
between stripes $D$.

In a similar manner to the case of the droplet phase, when
calculating the HF energy one needs to take into account the Hartree
interaction between the infinite number of unit cells in the
systems. Moreover, the extended nature of the states along the
stripes means that it is necessary to consider also the Fock
exchange between different unit cells on the same stripe. A detailed
account of the way this is done is presented in Appendix
\ref{app:hartree}.
The AFM spin coupling energy per unit area for both diagonal and
bond-aligned stripe phases is
\begin{equation}
  E_{J,stripe}=-2J\left[1-2m\left(2-\frac{1}{W}\right)\right],
\end{equation}
where $m=x/x_{FM}$ is the fraction of FM regions in the system.

By comparing the energies of all the above mentioned phases, a phase
diagram is constructed, depicting the nature of the ground state as
a function of the external parameters, $x$, $J/t$ and $U/t$.

\subsection{Monte-Carlo Simulated Annealing}

We have supplemented the calculation of the HF energy of various
variational configurations by an \emph{unrestricted} HF calculation
using Monte-Carlo simulated annealing. In this method, the energy of
a finite-sized system is minimized with respect to the full
configuration space of core spins, rather than a special subset of
spin textures. In each Monte-Carlo step the energy of a given
configuration of classical core spins is evaluated using the HF
approximation. A spin configuration is accepted as the system's new
state if the change in energy from the current state satisfies the
Metropolis condition. The temperature is slowly decreased until the
system reaches a stable, low energy configuration. If the
temperature is decreased slowly enough, the final state is the HF
approximation of the ground state.

The underlying assumption of the present study is that the system
indeed separates into FM and AFM regions with an abrupt boundary
between them. Therefore, the MC simulation needs to explore only
such configurations, improving the convergence time. This can be
achieved by setting all the spins on one sub-lattice to the ``up''
state, and incrementally flipping the spins on the other
sub-lattice. An additional improvement comes from a new algorithm
used to decide which spin to flip.  At first, a spin is chosen
randomly. It is flipped if the resulting state satisfies the
Metropolis criterion. If the spin is located near a FM-AFM boundary
then its neighbors are added to a queue of spins to be tested for
flipping. After all the spins in the queue have been tested for a
flip, a new spin is chosen randomly. Requiring that a spin be added
to the queue no more than once, prevents the simulation from
repeating itself, thus maintaining ergodicity.

In our calculations, the system contained $24\times24$ sites
arranged periodically on a torus. A linear annealing schedule was
employed over $50-100$ MC sweeps, and an identical number of sweeps
at the lowest temperature allowed the system to thermalize into the
ground state. The temperatures started from above $1.5J$ at the
beginning of the annealing schedule to below $0.5J$ at its end.

Even though this method minimizes the system's energy with respect
to an unrestricted configuration space, it has a number of
disadvantages when compared to the variational HF method, applied to
only a number of special configurations. First, its periodicity is
fixed; in our case it is $24$ sites along each axis. In addition,
the presence of long range interactions causes the simulations to
converge very slowly. Nevertheless, it provides an important
reference point with which the the variational HF results may be
contrasted, especially in order to confirm that the variational
manifold contains the most relevant configurations.

%%\subsection{Thermodynamic Equlibrium Considerations}

\section{Results}
\label{sec:results}

\begin{figure}[t]
  \centering
  \includegraphics[width=\linewidth]{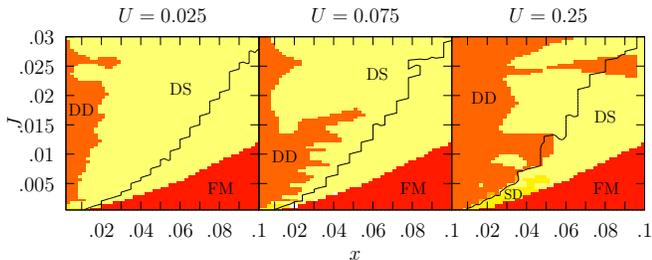}
  \caption{The HF phase diagram: DD - A triangular lattice of
    diamond-shaped droplets, SD - a square lattice of square droplets,
    DS - diagonal stripes, FM - a uniform ferromagnet. The black lines
    correspond to $x/x_{FM}=1/2$.}
  \label{fig:phsdgm}
\end{figure}

In the present section, we present and analyze our numerical results.
The coupling constants $J$ and $U$ are measured in units of $t$, by
setting the hopping amplitude $t=1$.  The HF energies of all the
considered phases were calculated in the parameter range $x\le 0.1$,
$J\le0.03$, and for three values of $U$, namely
$U=0.025,0.075,0.25$. We chose to concentrate on this region in the
$x-J$ plane for two reasons. As already mentioned, a previous
estimate\cite{jap02} sets $J^*=0.036$ as the upper limit for the
realization of a FM -- N\'{e}el AFM (as opposed to other types of
magnetic ordering) phase-separated state in two-dimensions.  Secondly,
our calculations indicate that the line $x/x_{FM}=1/2$, crosses
$J=0.03$ at $x^*=0.1$, see Fig. \ref{fig:phsdgm}. The region below
this line in the $x-J$ plane corresponds to configurations in which
the FM phase occupies more than half the system area.  While the
stripes phases, which we consider, continue to be relevant in this
region, we expect (and confirm in our MC simulations) that the phases
of FM droplets ought to be replaced by configurations of AFM droplets
embedded in a FM background.  The latter turn out to be more involved
computationally and were left out of the present study. We also wish
to note that the above values of $J^*$ and $x^*$, are sensitive to the
details of the considered model. Therefore, while experimentally,
percolation of the metallic phase at low temperatures is observed in
manganates with $x \agt 0.18$, we expect our qualitative conclusions
to apply to more complicated models of manganates, as long as phase
separation into FM and AFM phases is possible.

We begin our review of the results by discussing the phase diagram and
presenting general arguments for its structure. We then move on to
consider the details of the most dominant phases.

\subsection{The Phase Diagram}

The main result of our calculation is the phase diagram,
Fig. \ref{fig:phsdgm}, derived from the variational HF approach and
depicting the system's ground state configuration as a function of the
parameters, $x$, $J$ and $U$.  It demonstrates that diamond droplets
in a triangular formation is the preferred phase at low densities,
while diagonal stripes are prevalent at higher values of $x$.
%% GW added in response to the referee's report:
%%The stability of a striped phase is the most important result of our
%%calculation, and may have experimental implications, as we suggest
%%below.
%%
%Edited by DG:
This stability of a striped phase is the most important result of our
calculation. The striped arrangement is expected to possess unusual and
potentially useful properties (see Sec. \ref{sec:conclu} below, where we
also mention possible directions of experimental search for the stripe phase
in CMR manganates).
When the
Coulomb interaction strength $U$ is increased, the transition between
droplets and stripes occurs at higher $x$. As discussed above, our
variational approach becomes insufficient below the line
$x/x_{FM}=1/2$, as we do not allow for a phase of AFM droplets
embedded in a FM background. Such a phase is expected to appear near
the transition to the uniform FM state. This conclusion is supported
by the unconstrained HF results presented below. We are unable,
though, to map in detail the boundary between the stripe and droplet
phases in this parameter regime.

The general features of the phase diagram can be explained by simple
energy considerations. The preference of diagonal stripes and diamond
droplets is a direct result of the lower energy of diagonal
boundaries, as previously established by two of the
authors\cite{prb06}. The appearance of a triangular droplet lattice at
low densities is akin to the physics giving rise to the Wigner crystal
in a dilute gas of electrons. Next, we elaborate on the reasons and
nature of the transition between the droplet and stripe phases.

To this end, let us examine how the energy difference between the two
phases evolves with $x$. As $x$ increases, the distance between
droplets or stripes diminishes, but our HF results indicate that the
size of these FM regions and the electron density within them,
$x_{FM}$, vary slowly in the vicinity of the phase transition. The
combined difference between the kinetic and magnetic energies per
electron of the two phases, $\Delta\varepsilon$, depends on $x_{FM}$
and on the size and shape of the FM regions, but not on the distance
between them. We therefore conclude that at the qualitative level,
changes of $\Delta\varepsilon$ with $x$ cannot be the driving force
behind the transition. Instead, we concentrate on the doping
dependence of the difference in the Coulomb energy per electron
between droplets and stripes, $\Delta\phi$.

The Coulomb energy contains contributions coming from the interaction
between charges within a single super-lattice unit cell and between
different cells.  The neutrality of each unit cell (due to the
positive background) implies that the dominant contribution to the
Coulomb energy per electron originates from the intra-cell
component. Simple dimensional analysis allows us to obtain an estimate
for its behavior.
%Dimensional analysis of the Coulomb potential in the center of a
%droplet (stripe), due to other electrons in the same droplet (stripe)
%and the neutralizing background in the corresponding unit cell, can
%sufficiently describe these Coulomb driven transitions.
The amount of positive background charge within a droplet unit cell is
$xa^2$, $a$ being the inter-droplet spacing. Thus, the Coulomb potential
due to the positive background is $\phi_{droplet}^+ \approx -U x a$.
The interaction between electrons within a droplet generates
$\phi_{droplet}^-\approx Ux_{FM}L$, where $L$ is the droplet size.
Since $m\equiv x/x_{FM}\approx L^2/a^2$, we have
\begin{equation}
  \phi_{droplet} = \phi_{droplet}^- + \phi_{droplet}^+ \approx
  U x_{FM} L \left( 1 - \sqrt{m} \right).
\end{equation}
The Coulomb energy in the stripe phase takes a different form. The
amount of charge per unit length is $x_{FM}W = xD$ where $W$ and $D$ are the
stripe width and the distance between stripes, correspondingly. The background
potential is then $\phi_{stripe}^+ \approx Ux_{FM}W\ln D$ and the potential due
to electrons in the same stripe is $\phi_{stripe}^- \approx -Ux_{FM}W\ln W$.
Together they give
\begin{equation}
  \phi_{stripe} = \phi_{stripe}^- + \phi_{stripe}^+ \approx -Ux_{FM}W\ln m,
\end{equation}
where in this case $m=W/D$. Consequently, the difference in Coulomb energy per
electron between the droplet and stripe phases has the form
\begin{eqnarray}
  \Delta\phi & \equiv & \phi_{droplet}-\phi_{stripe} \nonumber \\
  & \approx &
  U x_{FM}\left[K_d L\left(1-\sqrt{m}\right) + K_s W \ln m \right],
\label{dphi}
\end{eqnarray}
\begin{figure}[ht]
  \centering
  \includegraphics[width=\linewidth]{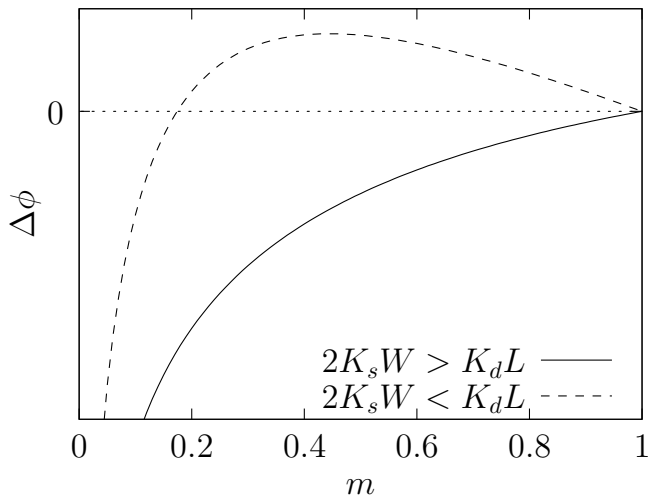}
  \caption{Dimensional analysis form of $\Delta\phi$ as a function of
    $m$, demonstrating the difference between the two possible
    branches.}
  \label{fig:dphith}
\end{figure}
\begin{figure}[ht]
  \centering
  \includegraphics[width=\linewidth]{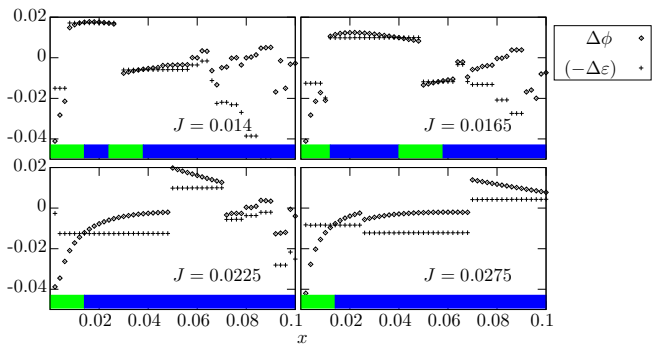}
  \caption{$\Delta\phi$ and $-\Delta\varepsilon$ as a function of $x$
    as obtained from the HF calculation for $U = 0.075$ and various
    values of $J$.  Below each plot are colored bands showing the
    ground state configuration at the corresponding $x$ values: green
    (light)- droplets, blue (dark) - stripes. The transitions occur
    when $\Delta\phi=-\Delta\varepsilon$.}
  \label{fig:dphi}
\end{figure}
where $K_d$ and $K_s$ are numerical constants characterizing the
geometry of droplets and stripes, respectively. It is implicitly
assumed in Eq. (\ref{dphi}) that in the transition region between the
phases $x_{FM}$ is the same for both configurations (the HF
calculation shows that this is correct up to $20\%$). The transition
itself takes place at $m^*$, satisfying
\begin{equation}
\Delta\phi(m^*)+\Delta\varepsilon=0,
\label{eq:transition}
\end{equation}
where we have used the constancy of $\Delta\varepsilon$ near the transition.
% i.e.,
%\begin{equation}
%  U x_{FM}\left[K_d\left(1-\sqrt{m^*}\right) + K_s\ln m^* \right]
%  = -\Delta\varepsilon
%\label{transcond}
%\end{equation}

If $2K_sW>K_dL$ then $\Delta\phi$ is a monotonously increasing
function of $m$ (in the physical range $0<m<1$), see
Fig. \ref{fig:dphith}. In this case at most a single solution, $m^*$,
exists to condition (\ref{eq:transition}), implying that the droplet
phase is preferred when $m<m^*$, while stripes occur for $m>m^*$;
% GW added this in order to make trends clear:
the area of the droplet phase increases with $U$.
On
the other hand, if $2K_sW<K_dL$, $\Delta\phi$ %changes
acquires a
maximum and two solutions, $m^*_1$ and $m^*_2$, may appear. Under such
conditions a reentrant behavior follows, i.e., droplets are preferred
when $m<m^*_1$ \emph{or} $m>m^*_2$, and stripes are realized in the
region $m^*_1<m<m^*_2$, which grows with increasing $U$. We note that
in any case the existence of a solution to condition
(\ref{eq:transition}), crucially depends on the value of
$\Delta\varepsilon$. It is the latter which reflects the features
taken into account for the first time in the present work ({\it viz.},
the orientational dependence of the boundary energy and the
quantization of the carrier motion.)

Fig. \ref{fig:dphi} shows HF results for $\Delta\phi$ and
$(-\Delta\varepsilon)$ as a function of $x$ for various values of
$J$. Transitions between droplet and stripe phases occur when
$\Delta\phi=-\Delta\varepsilon$. The division into discontinuous
segments is due to changes in the properties of the stripes or
droplets respectively (the optimal values of $W$, $L$, and $x_{FM}$,
see below). However, the transitions generally do not occur at these
points of discontinuity, leading to our previous assertions concerning
the constancy of $\Delta\varepsilon$ and the dominant role of the
Coulomb interaction in the vicinity of the transition. The first two
transitions at $J=0.014$ and $J=0.0165$ are near a maximum in
$\Delta\phi$, demonstrating the $2K_sW<K_dL$ branch behavior. Whereas
these transitions occur on one continuous segment, the third
transition occurs on a different segment, where only one solution
exists. Other one solution transitions are shown for $J=0.0225$ and
$J=0.0275$.

\subsection{The Droplet Phase}

\begin{figure}[ht]
  \centering
  \includegraphics[width=\linewidth]{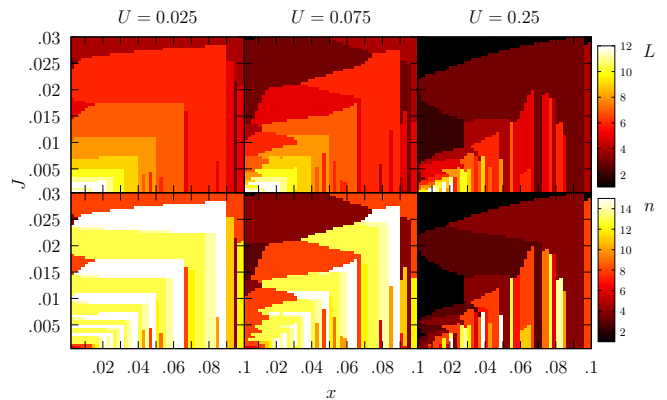}
  \caption{HF results for a triangular lattice of diamond droplets:
    optimal size $L$ (top) and number of electrons per droplet $n$
    (bottom).}
  \label{fig:ditr}
\end{figure}

In general, a triangular lattice of diamond shaped droplets proved to be
energetically more favorable than the other types of droplet phases.
As noted before this is a consequence of the directional dependence of the
FM-AFM boundary energy and the minimization of the inter-droplet Coulomb energy.
%Two reasons account for this result, as was
%briefly mentioned above. The reason diamond shaped droplets have a
%lower energy than square droplets is because diagonal boundaries have
%lower energy than boundaries parallel to the lattice, as shown in a
%previous paper.  A triangular formation of the droplets is expected to
%be favorable since, in the low density limit, it coincides with the
%expected Wigner crystal.
Fig. \ref{fig:ditr} shows the optimal droplet size, $L$, and the
number of conduction electrons per droplet, $n$, as deduced from the
variational HF calculation. Increasing the strength of the Coulomb
repulsion has the obvious effect of decreasing the droplet
size. Specifically, for the case of $U=0.25$ the variational study
yields very small ($L=1,2$) singly occupied droplets in the regime of
low $x$ and intermediate to large $J$. Comparing their energy to the
other types of inhomogeneous states reveals that these {\it magnetic
  polarons} are in fact the lowest energy configuration in this region
of parameters, see the HF phase diagram, Fig. \ref{fig:phsdgm}.

\subsection{The Stripe Phase}

\begin{figure}[ht]
  \centering
  \includegraphics[width=\linewidth]{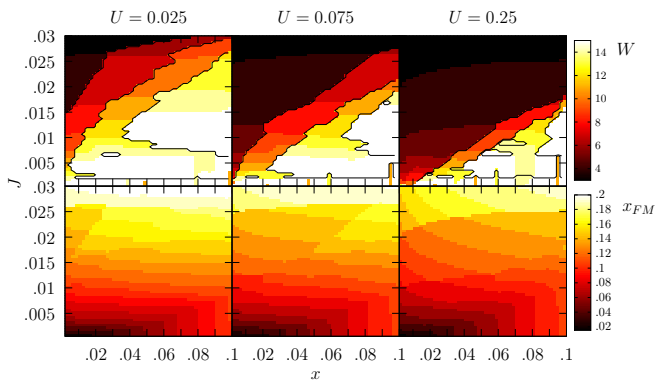}
  \caption{ HF results for diagonal stripes: optimal $W$ (top) and
    $x_{FM}$ (bottom).  The number of partially filled bands within a
    stripe changes across each black contour on the top panel.}
  \label{fig:diastr}
\end{figure}

Fig.\ref{fig:diastr} shows the optimal stripe width $W$ and conduction
electron density $x_{FM}$ for diagonal stripes. The latter are more
favorable than their bond-aligned counterparts due to the orientation
dependence of the boundary energy. One striking feature in these HF
results is the existence of abrupt transitions in the stripe width. A
small increase in $x$ may lead to a discontinuous change in $W$. On
the other hand, increasing $J$ typically leads to changes in $W$ which
are less steep. The electron density within the FM stripes, $x_{FM}$,
varies, in general, very slowly with $x$, and increases with $J$.

\begin{figure}[h]
  \centering
  \includegraphics[width=\linewidth]{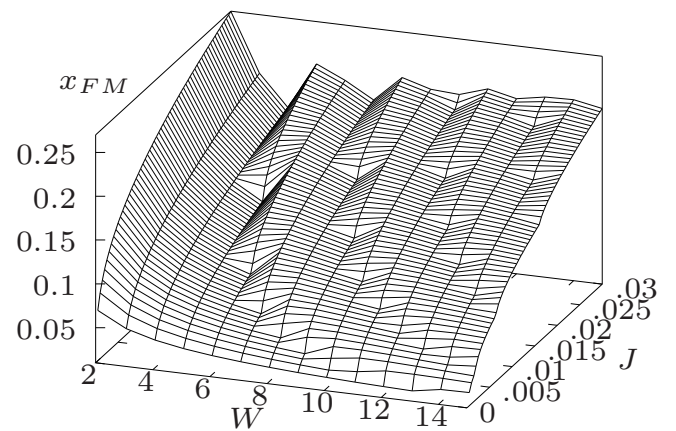}
  \caption{$x_{FM}$ in diagonal stripes of width $W$ as imposed by
    thermodynamic equilibrium.}
  \label{fig:xfmth}
\end{figure}

We use the condition of thermodynamic equilibrium,
Eq. (\ref{eq:tdstability}), between a diagonal FM stripe and its AFM
environment to explain these features.  The kinetic energy
contribution to the stripe's thermodynamic potential is determined by
its non-interacting spectrum consisting of $W$ bands (corresponding to
the quantization of transverse electron motion within the stripe)
\begin{equation}
  \varepsilon_b(k) = -t_b\cos\left(\frac{k}{2}\right),
\end{equation}
where $t_b$ is the bandwidth of band $b=1\cdots W$
\begin{equation}
  t_b = 2t \cos\left(\frac{b\pi}{W+1}\right).
\end{equation}
The resulting density of states in band $b$ is then
\begin{equation}
  g_b(\varepsilon)=\frac{2}{\pi t_b}\frac{1}{\sqrt{1-\left(\varepsilon/t_b
  \right)^2}},
  %=\frac{1}{2\pi}\left|\frac{d\varepsilon}{dk}\right|^{-1}
  %% \left|\pi t_b \sin \left(\frac{k}{2}\right)\right|^{-1}
  %\left|\pi t_b \sin
  % \cos^{-1}\left(-\frac{\varepsilon}{t_b} \right)\right|^{-1},
\end{equation}
which together with the chemical potential $\mu$ determines the number
of electrons $n_b$ per unit length in the band. Since we are
interested in relatively low doping levels we consider the lower $W/2$
bands for which
\begin{equation}
  n_b = \int_{-t_b}^\mu g_b(\varepsilon)d\varepsilon
  = \frac{2}{\pi}\cos^{-1}\left(-\frac{\mu}{t_b}\right),
\end{equation}
and the non-interacting electronic contribution to the total energy is
\begin{equation}
  E_b = \int_{-t_b}^\mu \varepsilon g_b(\varepsilon)d\varepsilon
  = -\frac{2t_b}{\pi}\sin\left(\frac{\pi n_b}{2}\right).
\end{equation}
Using these terms, the thermodynamic potential in the FM stripes is
\begin{equation}
  \Omega_{FM}(W) = \frac{1}{W}\sum_b(E_b-\mu n_b)
  +2J\left(1-\frac{2}{W}\right),
\end{equation}
where the second term is the magnetic energy, taking into account the
structure of the boundaries. $\Omega_{AFM}$ remains the same as for an
infinite AFM region, Eq. (\ref{eq:omegaafm}). Fig. \ref{fig:xfmth}
shows $x_{FM}=\sum_bn_b/W$ evaluated at the Fermi energy $\mu$ which
solves $\Omega_{FM}(W)=\Omega_{AFM}$ as a function of $J$ \emph{and}
$W$.

Points of non-analyticity occur whenever the chemical potential
increases beyond the bottom of a band, $\mu =-t_b$, so that the
carriers begin to fill this additional band.  These non-analyticities
result in a corrugated landscape for $x_{FM}(W,J)$, shown in
Fig. \ref{fig:xfmth}, whereby several values of $W$ may correspond to
the same $x_{FM}$.  Thus, a small increase in $x$ may drive an abrupt
change in $W$ but leave $x_{FM}$ constant.  This transition is
accompanied by a change in the number of partially filled bands within
the stripe, as depicted by the black contours in
Fig. \ref{fig:diastr}.

\subsection{Simulation Results}

\begin{figure}[ht]
  \includegraphics[width=0.48\linewidth]{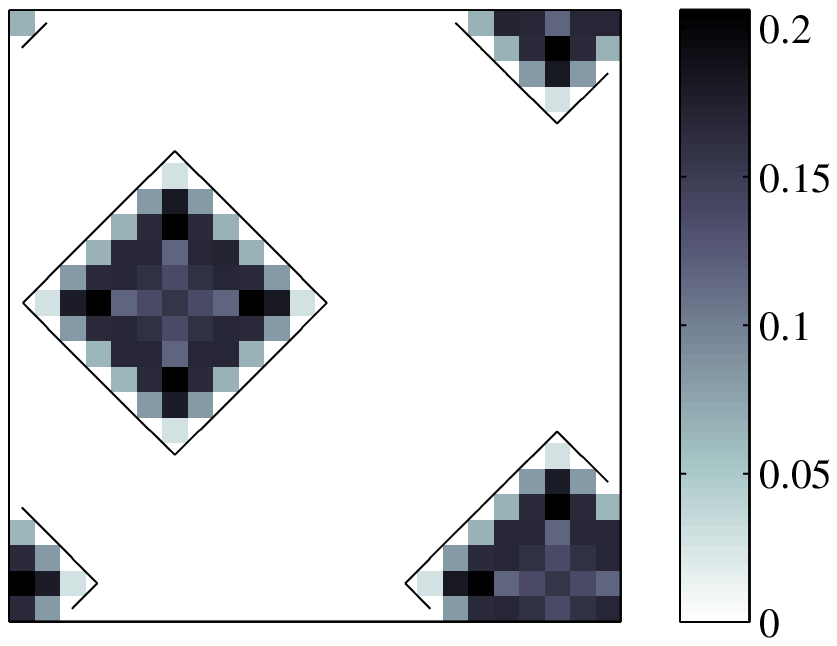}
  \includegraphics[width=0.48\linewidth]{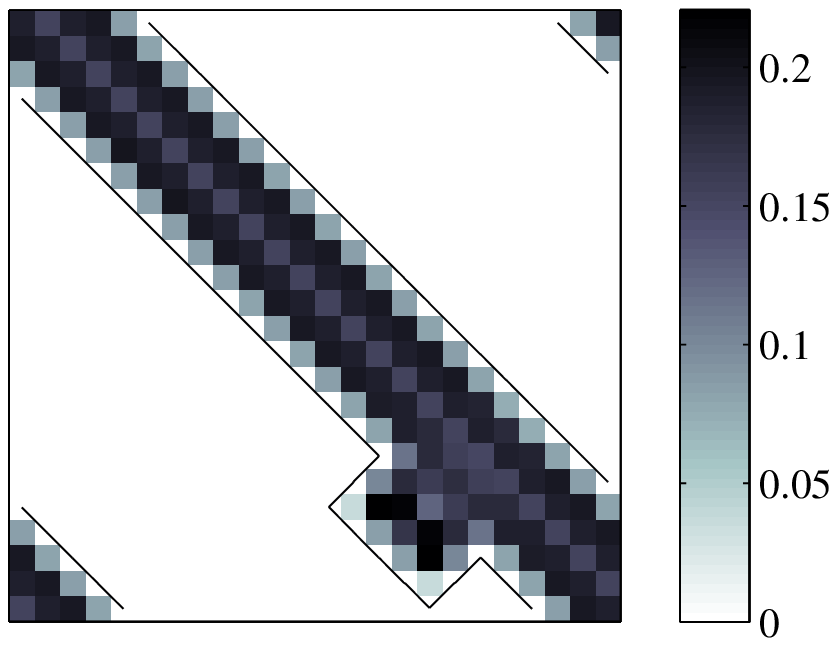}\\
  \includegraphics[width=0.48\linewidth]{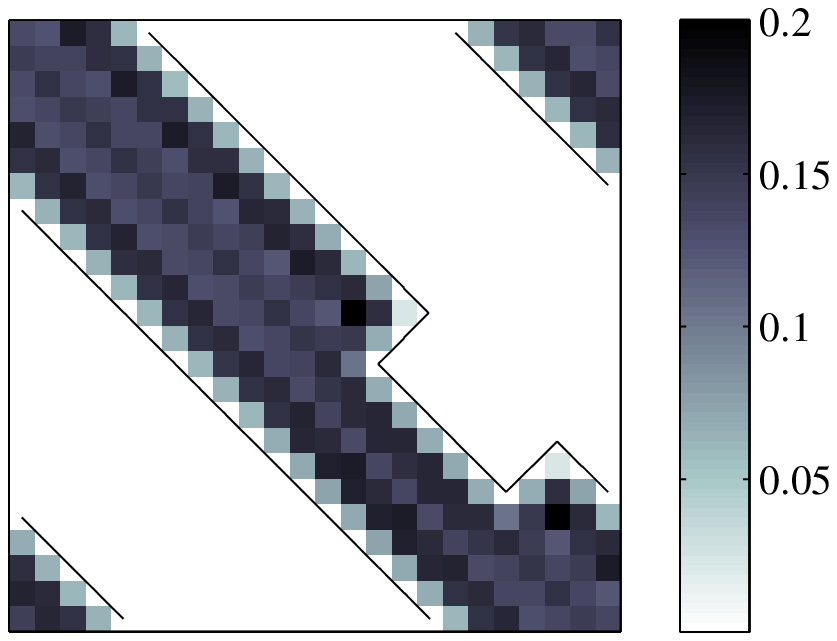}
  \includegraphics[width=0.48\linewidth]{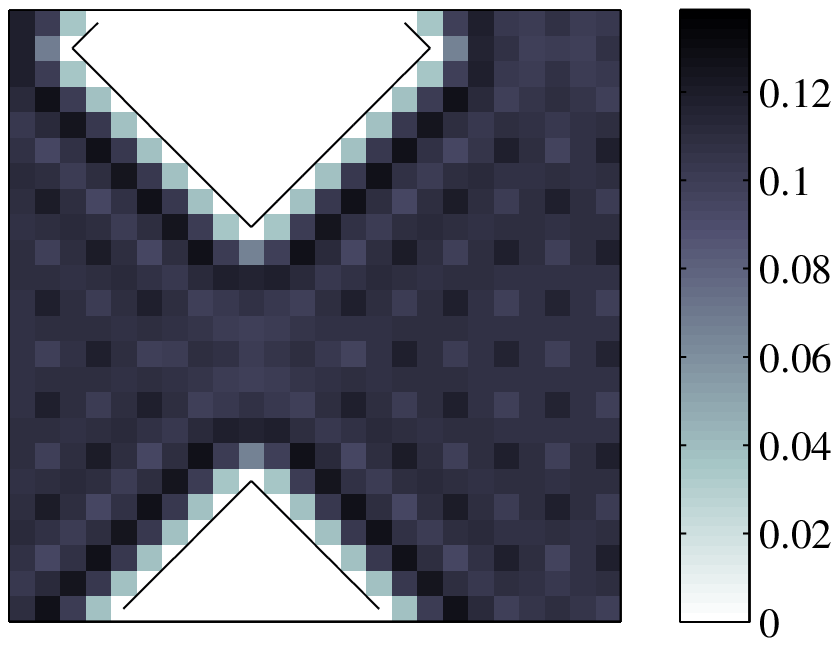}\\
  \caption{(color online) Electron density in ground states reached
    using simulated annealing.
    Top left: $x = 0.0278$, $J=0.015$, $U=0.075$.
    Top right: $x=0.0486$, $J=0.02$, $U=0.05$.
    Bottom left: $x = 0.0556$, $J=0.015$, $U=0.05$.
    Bottom right: $x = 0.0833$, $J=0.01$, $U=0.05$.
    Dark lines outline FM-AFM boundaries.}
  \label{fig:montecarlo}
\end{figure}

Although our Monte-Carlo results are not sufficient for constructing
the phase diagram, they yield convincing evidence that the phases
included in the variational HF calculation are indeed the appropriate
variational phases to consider. Some examples of ground states
obtained by MC simulated annealing are given in Fig.
\ref{fig:montecarlo}. Note that the moderate cluster size used in the
simulation induces finite-size effects apparent, for example, in the
imperfections of the stripe configurations. These results, in addition
to results from other simulations done at other parameter values,
agree with the general structure of the phase diagram in
Fig. \ref{fig:phsdgm}. Moreover, the unrestricted nature of the MC
method yields also configurations with AFM droplets in a FM background
(bottom right of Fig. \ref{fig:montecarlo}). As mentioned before, such
states were not considered in the variational HF approach because of
the relative difficulty in calculating their HF energy. Nevertheless,
there are reasons to believe, as is confirmed by the simulations, that
such a phase indeed exists around the transition line between the
striped and uniform FM phases, where $m=x/x_{FM}>1/2$.

\section{Discussion}
\label{sec:conclu}

The main finding of this paper concerns the geometry of the low
temperature phase-separated state in a two-dimensional
double-exchange magnet. We did not invoke any lattice or orbital
degrees of freedom but instead concentrated on the effects of the
ubiquitous long range Coulomb interaction. We verified that when the
relative area occupied by the FM phase, $x/x_{FM}$, is sufficiently
large, a {\it striped arrangement} (rather than a droplet
super-lattice) is stabilized. Our results also confirm the
expectation, based on a previous analysis of the directional
dependence of the FM-AFM boundary energy\cite{prb06}, that
diamond-shaped droplets and diagonal stripes are preferred over
their square and bond-aligned counterparts. Indications to this
effect are also present in other numerical studies of
double-exchange models\cite{DEnumerical}.

The stability of the stripe phase should not come as a surprise. In
fact, even in the earlier studies, which considered the continuum
limit\cite{Nagaevbook,Nagaev72}, it was noted that the energies of
the stripe and droplet configurations can be very close, although no
parameter window was found where stripes would correspond to the
lowest energy configuration. As a result, it is conceivable, as
indeed was shown in Ref. \onlinecite{Nagaev01}, that a stripe phase
may be stabilized, even in this limit, once physics due to some sort
of additional degrees of freedom is taken into account. Stripes also
occur naturally in other models, such as $t-J$ or
Hubbard\cite{Hubbard}, which involve a competition between the AFM
nature of a parent undoped state and the kinetic energy of the doped
charge carriers. Long-range AFM interaction was found
to favor stripes in the FM Ising model\cite{Low}.
Regarding the case of a pure double-exchange system with Coulomb
repulsion considered here, it has already been
argued\cite{prb06,prb03} that the correct treatment of the
boundaries between the FM and AFM regions is likely to tilt the
balance in favor of a striped arrangement.

% In the present work we were able to span the range between the
% macroscopic phase-separated state and the experimentally relevant
% case of nanometer scale FM inclusions comprising only few lattice
% periods, where quantization effects are essential.
In the present work, we considered the experimentaly relevant case
of nanometer-size FM inclusions (comprising only a few
lattice periods). Beside mapping the evolution of the geometry of
the inhomogeneous system, we addressed the long-standing question
regarding the stability of free magnetic polarons. As expected, we
find polaronic behavior in the region of small carrier concentration
$x\ll 1$ and strong Coulomb interaction. Away from this regime,
individual magnetic polarons coalesce into larger FM areas. We were
able to span the entire intermediate regime  between the coventional
phase separation (where the quantized character of the carrier motion
becomes unimportant) and an array of free magnetic polarons
(for which the notion of thermodynamic equilibrium between FM
and AFM phases becomes irrelevant). We emphasize that
the two main physical ingredients underlying our findings, namely,
the quantized electronic motion in small FM regions and the
directional dependence of the boundary energy, can be viewed as
largely model-independent. Therefore, our present conclusions can be
expected to stand for any double-exchange model with a long range
interaction, including the case when the latter originates from
crystal strain fields\cite{Ahn,Khomskii}.

The bulk of our study was carried out using a variational HF
approximation for the energy of various droplet and stripe phases.
It was supplemented by unconstrained HF calculations on moderate
size clusters, implemented via Monte-Carlo simulated annealing. The
HF approximation is expected to gain accuracy whenever the ratio of
electrostatic energy to kinetic energy is small. Throughout the
range of parameters studied by us, this ratio never exceeds 0.15.
Moreover, since we deal with the case of fully polarized electronic
spins, the spatial part of the many-body wave-function is
antisymmetric. This fact reduces correlation corrections to the HF
result which stem from the tendency of any pair of electrons, owing
to their mutual repulsion, to be more distant from each other than
the HF wave-function would indicate.

We close with a brief discussion of the experimental situation. To
the best of our knowledge, a conclusive experimental observation of
metallic stripes in phase-separated films of CMR materials
is yet to
be made. We note, however, that stripe-like charge ordering on the
atomic scale (charge density wave) was observed in a variety of
manganates. This includes films with different doping
levels\cite{stripesfilm}, as well as ceramic\cite{loudon}
 and single crystal\cite{Ronnow} samples in the
insulating state above $x=0.5$. In addition,
it was suggested\cite{filaments} that the phase separated state in a
three-dimensional system may acquire a filament structure.

Nevertheless, we argue that it would be desirable to synthesize
manganate films whose phase-separated state clearly exhibits
metallic stripes.
In addition to illustrating our theoretical picture, such
systems are expected to display unique and potentially useful
properties, some of which were not previously observed. One of these
is an anisotropic conductance, whereby the
stripes direction determines a low-resistivity axis, which ought to
be amenable to reorientation by, {\it e.g.}, applying a voltage. In
general, one expects to find history-dependent resistance and memory
effects akin to, and probably more pronounced than those observed
earlier in phase separated films\cite{phasepfilms,Vlakhov}, for which
not evidence for stripes was reported.
% These effects should be more pronounced in the stripy system but
% more pronounced since now, instead of individual conduction paths,
% the sample would have stripes running through its entire area.
When the sample composition gets close to the one corresponding to a
stable striped arrangement, weak perturbations such as external
electric or magnetic fields may be sufficient to change the geometry
of the FM regions from droplets to stripes, with a drastic change in
transport properties in the form of colossal electroresistance due
to dielectrophoresis\cite{phoresis} and large low-temperature
magnetoresistance.

Which manganate system could potentially exhibit a metallic stripe
order? In general, in order to look for such a state one is
interested to explore the parameter space by changing the average
carrier concentration $x$, the metallic area fraction $m$, and
the strength $U$ of the Coulomb interaction\cite{varyU}. While $x$
is determined by the dopant concentration, the experimentally
measurable quantity $m$ depends, in our model, on the ratio of
the AFM coupling $J$ to the hopping $t$. The latter may be affected
by, {\it e.g.,} the choice of the rare earth ion. An example of a
system which apparently allows control over the value of $m$ is
${(\rm La_{1-y} Pr_y)_{1-x} Ca_x Mn O_3}$. Three-dimensional
crystals of this compound with $x$ between 0.25 and 0.5, are
metallic for $y=0$, with no signatures of phase separation at low
temperatures\cite{Schiffer}. At $y=1$, the system is
phase-separated\cite{PCMO}, and exhibits robust insulating behavior,
presumably corresponding to well-separated metallic droplets in an
insulating matrix. The properties of the phase-separated state
change as one decreases the value of $y$, and at $y=0.7$ it is
possible to observe conduction paths formation and switching as a
result of an applied current\cite{Tokunaga}. Similar behavior is
also found in thin films of the same compound, which at least for
sufficiently large values of $y$ are
phase-separated\cite{LPCMOfilms}, as reflected in their peculiar
dielectric and transport properties\cite{Zhai,Guneeta}.

These findings prompt us to suggest looking for signatures of stripes
in ${(\rm La_{1-y} Pr_y)_{1-x} Ca_x Mn O_3}$ films by systematically
varying $y$, and with it, as indicated above, the relative area $m$ of
the metallic phase.  We expect stripes to appear around the point
where the areas of metallic an insulating phases are equal to each
other.  Besides ${(\rm La_{1-y} Pr_y)_{1-x} Ca_x Mn O_3}$, there are
other hole-doped manganate systems which may exhibit a stripe geometry
of phase separation, see Ref. \onlinecite{PCSMO}. In addition, we
expect our results to be relevant for some electron-doped
manganates\cite{eldoped}, as well as possibly for Eu-based magnetic
semiconductors\cite{Nagaevbook}.

\acknowledgments{The authors take pleasure in thanking R. Berkovits,
  J. T. Chalker, and G. Singh-Bhalla for enlightening discussions.
  This work was supported by the United States - Israel Binational
  Science Foundation (grant No. 2004162). Support by the Israeli
  Absorption Ministry is also acknowledged.}

\appendix
\section{Hartree--Fock Equations for a Periodic Configuration}
\label{app:hartree}

The HF equations for $n$ interacting, spin polarized,
electrons, may be written in matrix form as an effective eigenvalue
equation, which needs to be solved self-consistently\cite{hartree}
\begin{equation}
  \sum_{\bf r'}H_{\bf rr'}\phi_{s\bf r'}=\varepsilon_{s}\phi_{s\bf r},
  \label{eq:EV1}
\end{equation}
where the effective Hamiltonian matrix is given by
\begin{equation}
  H_{\bf rr'}=h_{\bf rr'}+\sum_{s=1}^{n}\left(\delta_{\bf rr'}\sum_{\bf r''}v_{\bf rr''}
    \left|\phi_{s\bf r''}\right|^{2}-v_{\bf rr'}\phi_{s\bf r'}^{*}\phi_{s\bf r}\right).
    \label{HHF}
\end{equation}
The indices ${\bf r}$, ${\bf r'}$ and ${\bf r''}$ indicate positions
on the lattice, $h_{\bf rr'}$ is the single-particle part of the
Hamiltonian, and $v_{\bf rr'}$ is the interaction energy of a particle
at site $\bf r$ and a particle at site $\bf r'$. $\phi_{s\bf r}$ are
the eigenvectors in lattice-site representation, each indexed by label
$s$ and with $\varepsilon_s$ as its eigenvalue. The self-consistent
solution yields the HF ground state energy, given by
\begin{equation}
  E=\frac{1}{2}\sum_{s=1}^{n}\left(\varepsilon_{s}+\sum_{\bf rr'}\phi_{s\bf r}^{*}
  h_{\bf rr'}\phi_{s\bf r'}\right),
\end{equation}
where the summation is over the $n$ states with lowest eigenvalues
$\varepsilon_s$. For the Hamiltonian, Eq. (\ref{eq:Ham_simp}),
considered in the present study the single-particle term is
\begin{equation}
  h_{\bf rr'}=\frac{t_{\bf rr'}}{2}-\delta_{\bf rr'}Ux\int\frac{d\mathbf{R}}
  {\left|\mathbf{r}-\mathbf{R}\right|},
  \label{HFsinglepart}
\end{equation}
with an implicit dependence, given by Eq. (\ref{teff}), of $t_{\bf
  rr'}$ on the configuration of the core spins $\left\{ S_{\bf
    r}\right\} $. The second term in Eq. (\ref{HFsinglepart}) reflects
the interaction between the conduction electrons and a {\it
  continuous} neutralizing positive background of density $x=n/A$,
where $A$ is the system area, via the Coulomb potential $v_{\bf
  rr'}=U/\left|\mathbf{r}-\mathbf{r}'\right|$.
%and a constant Coulomb
%interaction with a uniform positively charged background (we consider
%a \emph{continuous} background charge distribution for simplicity).
%This infinite background term is needed in order to counter the
%infinite repulsion energy of the electrons. In addition, the
%interaction of the background with itself will be
%included. Nevertheless, we shall see that all these infinite terms
%cancel each other. The Coulomb interaction energy between two
%conduction electrons is just
%%\begin{equation}
%% v_{rr'}=\frac{U}{\left|\mathbf{r}-\mathbf{r}'\right|}.
%%\end{equation}
%$v_{rr'}=U/\left|\mathbf{r}-\mathbf{r}'\right|$.
%Since the background charge density is equal to the average electron
%density, $x=n/V$, where $V$ is the system volume (or area in the 2D case).
Noting that the eigenvectors are normalized to unity, $\sum_{\bf
  r''}\left|\phi_{s{\bf r''}}\right|^{2}=1$, and that the ${\bf
  r}={\bf r''}$ and ${\bf r}={\bf r'}$ terms in Eq. (\ref{HHF}) are
equal and opposite, we are led to analyze the following HF Hamiltonian
\begin{widetext}
\begin{equation}
  H_{\bf rr'}=\frac{t_{\bf rr'}}{2}+U\sum_{s=1}^{n}\left[\delta_{\bf rr'}\sum_{\bf r''}
      \left( \frac{1-\delta_{\bf rr''}}{\left|\mathbf{r}-\mathbf{r''}\right|}
      -\frac{1}{A}\int\frac{d\mathbf{R}}
      {\left|\mathbf{r}-\mathbf{R}\right|}\right)\left|\phi_{s{\bf r''}}\right|^{2}
    -\frac{1-\delta_{\bf rr'}}{\left|\mathbf{r}-\mathbf{r'}\right|}
    \phi_{s{\bf r'}}^{*}\phi_{s{\bf r}}\right].
  \label{eq:fullmatrix}
\end{equation}
\end{widetext}

We are interested in cases where the core spins configuration is
periodic, such that the system can be divided into $N$ unit cells,
each containing an identical configuration of spins on $M$ sites. Let
the super-lattice vectors $\left\{ \mathbf{l}\right\} $ identify the
location of the unit cells. A position ${\bf r}$ on the lattice can
then be written as $
\mathbf{r}\left(\mathbf{l},i\right)=\mathbf{l}+\mathbf{r}_{i}$, where
${\bf r}_i$ is the position within the unit cell ${\bf l}$, containing
${\bf r}$.  The spin periodicity implies that $H_{\bf rr'}$ between
sites $\mathbf{r}=\mathbf{l}+\mathbf{r}_{i}$ and
$\mathbf{r'}=\mathbf{l'}+\mathbf{r}_{j}$ depends only on $i$, $j$, and
the super-lattice vector ${\bf l''}={\bf l'}-{\bf l}$ connecting the
two unit cells, {\it i.e.}, $H_{\bf rr'}=H_{ij}({\bf l''})$. As a
consequence of Bloch's theorem this means that the energy
eigenfunctions, expressed in the $({\bf l},i)$ representation, take
the form
$\phi_{bi}(\mathbf{k})e^{i\mathbf{k}\cdot\mathbf{l}}/\sqrt{N}$, with
eigenenergies $\varepsilon_b({\bf k})$, where $\mathbf{k}$ is defined
within the first Brillouin zone of the reciprocal super-lattice. The
``band'' index $b$ runs from $1$ to $M$, and $\phi_{bi}(\mathbf{k})$
is normalized to unity within a single unit cell. Written in the
$({\bf k},i)$ basis, the Hamiltonian becomes block diagonal, where the
matrix elements of the block connecting states with the same ${\bf k}$
are given by
\begin{widetext}
\begin{equation}
  H_{ij}\left(\mathbf{k}\right)=\frac{t_{ij}\left(\mathbf{k}\right)}{2}
  +\frac{U}{N}\sum_{b,\mathbf{k}'}
  \Theta\left[\mu-\varepsilon_{b}(\mathbf{k}')\right]
  \left[\delta_{ij}\sum_{i'}v_{ii'}^{H}
    \left|\phi_{bi'}\left(\mathbf{k}'\right)\right|^{2}
    -v_{ij}\left(\mathbf{k}-\mathbf{k'}\right)
    \phi_{bj}^{*}\left(\mathbf{k}'\right)
    \phi_{bi}\left(\mathbf{k}'\right)\right].
    \label{HFHk}
\end{equation}
\end{widetext}
Here
\begin{equation}
t_{ij}\left(\mathbf{k}\right)=\sum_{\mathbf{l}}t_{ij}
\left(\mathbf{l}\right)e^{-i\mathbf{k}\cdot\mathbf{l}},
\label{eq:kinsum}
\end{equation}
is the Fourier transform of the hopping amplitudes
$t_{ij}(\mathbf{l})$ between sites $i$ and $j$ in unit cells separated
by a super-lattice vector $\mathbf{l}$. We also introduced
\begin{eqnarray}
  &&v_{ij}\left(\mathbf{k}\right) = \sum_{\mathbf{l}}
  \frac{1-\delta_{ij}\delta_{\mathbf{l},\mathbf{0}}}
  {\left|\mathbf{l}+\mathbf{r}_{i}-\mathbf{r}_{j}\right|}
  e^{-i\mathbf{k}\cdot\mathbf{l}},\label{eq:focksum}\\
  &&v_{ii'}^{H} = v_{ii'}\left(0\right)-\frac{1}{A_{u.c.}}
  \int\frac{d\mathbf{R}}{\left|\mathbf{r}_i-\mathbf{R}\right|},
  \label{eq:hartreesum}
\end{eqnarray}
where $A_{u.c.}=A/N$ is the area of a unit cell. The chemical potential, $\mu$,
is defined by $n=\sum_{b,\mathbf{k}}\Theta[\mu-\varepsilon_{b}(\mathbf{k})]$,
with $\Theta(x)$ denoting the step function.

The HF ground-state energy of the conduction electrons is
\begin{eqnarray}
\nonumber
E_{el} &=& \frac{1}{2}\sum_{b,\mathbf{k}}
  \left[\varepsilon_{b}\left(\mathbf{k}\right)
    +\frac{1}{2}\sum_{ij}\phi_{bi}^{*}\left(\mathbf{k}\right)
    t_{ij}\left(\mathbf{k}\right)\phi_{bj}\left(\mathbf{k}\right)\right]\\
&\times & \Theta\left[\mu-\varepsilon_{b}(\mathbf{k})\right]
 -\frac{1}{2}Unx\int\frac{d\mathbf{R}}{\left|\mathbf{R}\right|},
  \label{eq:fullenergy}
\end{eqnarray}
where in the last term we have taken the limit
$N\rightarrow\infty$. This diverging contribution is canceled by the
self-interaction of the positive background, evaluated in the same
limit
\begin{equation}
  E_{bg}=\frac{1}{2}Ux^{2}\int\frac{d\mathbf{R}\, d\mathbf{R}'}
  {\left|\mathbf{R}-\mathbf{R}'\right|}
  =\frac{1}{2}Unx\int\frac{d\mathbf{R}}{\left|\mathbf{R}\right|}.
\end{equation}
Consequently, the total energy (not including the contribution of the
anti-ferromagnetic interaction between the core spins) is given by the
sum over $b$ and ${\bf k}$ in Eq. (\ref{eq:fullenergy}).

In order to evaluate the matrix elements of the HF Hamiltonian, we
need a method to calculate the infinite super-lattice sums in
Eqs. (\ref{eq:kinsum})-(\ref{eq:hartreesum}).  The first of these is
trivial since hopping is allowed only between nearest-neighbor sites
within a unit cell or between adjacent ones. In two dimensions this
leaves at most five terms to the sum. On the other hand, the Coulomb
interaction is long-ranged, and an infinite number of terms needs to
be included in Eqs. (\ref{eq:focksum}) and (\ref{eq:hartreesum}).

\subsection{The Hartree term}

The Hartree interaction matrix (\ref{eq:hartreesum}) includes two
diverging contributions, one coming from the interaction with the
average electronic density and the other from the interaction with the
positive uniform background. The two contributions cancel each
other. In order to demonstrate this and extract the remaining finite
piece we employ Ewald summation (see Appendix \ref{app:ewald}). The
main identity of this method, directly applicable to the evaluation of
the first term in Eq. (\ref{eq:hartreesum}), is
\begin{eqnarray}
  \sum_{\mathbf{l}}\frac{1}{\left|\mathbf{l}+\mathbf{r}\right|} & = &
  \frac{2\pi}{A_{u.c.}}\sum_{\mathbf{g}}\frac{e^{i\mathbf{g}\cdot\mathbf{r}}}
  {\left|\mathbf{g}\right|}\mbox{erfc}\left(\frac{\left|\mathbf{g}\right|}
    {2G}\right) \nonumber \\
  & & +\sum_{\mathbf{l}}\frac{1}{\left|\mathbf{l}+\mathbf{r}\right|}
  \mbox{erfc}\left(G\left|\mathbf{l}+\mathbf{r}\right|\right).
  \label{eq:ewaldquote}
\end{eqnarray}
As before, $\mathbf{l}$ are the super-lattice vectors and $A_{u.c.}$
is the unit cell area. Here, $\mathbf{g}$ are the reciprocal
super-lattice vectors, and $G$ is an arbitrary constant, chosen to
minimize the number of relevant terms in both sums controlled by the
complementary error function $\mbox{erfc}\left(x\right)$. Note that
the divergence which stems from summing over large ${\bf l}$ vectors
in the left hand side of Eq. (\ref{eq:ewaldquote}) is encoded in the
$\mathbf{g}=0$ term on the right hand side. This divergence is
canceled by the integral over the whole system in
Eq. (\ref{eq:hartreesum}). This can be readily seen by using
Eq. (\ref{eq:ewaldquote}) with $G\to\infty$, to write it as
\begin{equation}
  \frac{1}{A_{u.c.}}\int\frac{d\mathbf{r}}{\left|\mathbf{r}\right|} =
  \frac{1}{A_{u.c.}}\int_{u.c.}d\mathbf{r}\sum_{\mathbf{l}}
  \frac{1}{\left|\mathbf{l}+\mathbf{r}\right|}
%%  & = & \frac{2\pi}{A^{2}}\sum_{\mathbf{g}}\frac{1}{\left|\mathbf{g}\right|}
%%  \int_{u.c.}d^{2}r\, e^{i\mathbf{g}\cdot\mathbf{r}} \nonumber \\
  =  \frac{2\pi}{A_{u.c.}}\sum_{\mathbf{g}}\frac{\delta_{\mathbf{g},\mathbf{0}}}
  {\left|\mathbf{g}\right|}.
\end{equation}
Consequently we find for the Hartree matrix
\begin{eqnarray}    v_{ii'}^{H} & = &\, \frac{2\pi}{A_{u.c.}}
  \sum_{\mathbf{g}\neq\mathbf{0}}
    \frac{e^{i\mathbf{g}\cdot\mathbf{r}_{ii'}}}{\left|\mathbf{g}\right|}\mbox{erfc}
    \left(\frac{\left|\mathbf{g}\right|}{2G}\right)
    -\frac{2\sqrt{\pi}}{A_{u.c.}G} \nonumber \\
    & & +\sum_{\mathbf{l}\neq\mathbf{0}}\frac{1}{\left|\mathbf{l}
        +\mathbf{r}_{ii'}\right|}\mbox{erfc}\left(G\left|\mathbf{l}
        +\mathbf{r}_{ii'}\right|\right) \nonumber \\
    & & +\left(1-\delta_{ii'}\right)
    \frac{\mbox{erfc}\left(G\left|\mathbf{r}_{ii'}\right|\right)}
    {\left|\mathbf{r}_{ii'}\right|}-\delta_{ii'}\frac{2G}{\sqrt{\pi}},
\end{eqnarray}
where ${\bf r}_{ii'}={\bf r}_i-{\bf r}_{i'}$.

\subsection{The droplets Fock term}

When the core spins are arranged in FM droplets separated by an AFM
ordered background, the conduction electrons cannot hop from one unit
cell to the other, {\it i.e.},
$t_{ij}\left(\mathbf{l}\right)=t_{ij}\delta_{\mathbf{l},\mathbf{0}}$.
Consequently $t_{ij}\left(\mathbf{k}\right)$ is independent of ${\bf
  k}$, see Eq. (\ref{eq:kinsum}). Under such a condition it is easy to
verify that the HF eigenfunctions and eigenenergies are ${\bf
  k}$-independent as well. To prove this assertion, let us assume that
it is true and show that it leads to a ${\bf k}$-independent HF
Hamiltonian, hence closing the argument self-consistently.  Since the
Hartree term in the HF Hamiltonian, Eq. (\ref{HFHk}), depends on
$\mathbf{k}$ only through the HF eigenfunctions it obviously fulfills
the requirement. To complete the demonstration we note that the same
is true for the Fock term since it satisfies
\begin{eqnarray}
  H_{ij}^{Fock}\left(\mathbf{k}\right) & = & -\frac{U}{N}
  \sum_{b,\mathbf{k}'}\Theta(\mu-\varepsilon_{b})
  v_{ij}\left(\mathbf{k}-\mathbf{k}'\right)\phi_{bi}^{*}\phi_{bj},
  \nonumber\\
  & = & -U\sum_{b}\Theta(\mu-\varepsilon_{b})\frac{1-\delta_{ij}}
  {\left|\mathbf{r}_{i}-\mathbf{r}_{j}\right|}\phi_{bi}^{*}\phi_{bj}.
  \label{Fockdrop}
\end{eqnarray}
Moreover, Eq. (\ref{Fockdrop}) implies that in the case of FM droplets
the calculation of the Fock term involves only a finite sum (over the
$M$ states within each droplet). This is a direct consequence of the
vanishing overlap between electronic states in different droplets.

\subsection{The stripes Fock term}

When the core spins are arranged in a striped configuration, hopping
is allowed between unit cells along the direction of the stripes. In
other words, if we decompose the super-lattice vectors as
$\mathbf{l}=n_{a}\mathbf{a}+n_{b}\mathbf{b}$, where $\mathbf{a}$ and
$\mathbf{b}$ are primitive vectors along and off the stripe direction,
respectively, then
$t_{ij}\left(n_{a},n_{b}\right)=t_{ij}\left(n_{a}\right)\delta_{n_{b},0}$.
As a result $t_{ij}\left(\mathbf{k}\right)$, depends only on the ${\bf
  k}$ component along the stripes, {\it i.e.},
$t_{ij}\left(\mathbf{k}\right)=t_{ij}\left(k_a\right)$.  It follows
then, using the same reasoning presented above for the droplet case,
that the HF eigenfunctions and eigenenergies depend only on $k_a$, and
the Fock term takes the form
\begin{eqnarray}
  \nonumber
  H_{ij}^{Fock}\left(k_a \right) & = & -\frac{U}{N_a}
  \sum_{b,k_a'} \Theta[\mu-\varepsilon_{b}(k_a')]\phi_{bi}^{*}(k_a')\phi_{bj}(k_a') \\
  &\times &\sum_{n_a} \frac{1-\delta_{ij}\delta_{n_a,0}}
  {|{\bf r}_{ij}+n_a{\bf a}|}e^{in_a (k_a-k_a') a},
  \label{Fockstripe}
\end{eqnarray}
where $N_a$ is the number of unit cells along the stripe,
$a=|\mathbf{a}|$, and ${\bf r}_{ij}$ is the vector connecting sites
$i$ and $j$ within a unit cell.

In contrast to the Hartree term where the interaction decays slowly,
the exponential factor in the Fock exchange, Eq. (\ref{Fockstripe}),
ensures that the series converges relatively fast. Hence, the infinite
sum is well approximated by assuming a long but finite stripe. In our
calculation, we used $N_a=100-200$, and verified that larger values
change the ground state energy by an insignificant amount. Note that
the logarithmic divergence of the $n_a$ sum in the case $k_a'=k_a$, is
integrable, and vanishes upon the summation over $k_a'$.

\section{Ewald Summation in 2D}
\label{app:ewald}

The development (based on Ref. \onlinecite{ewalds}) of Ewald's
summation method begins with defining the function
\begin{equation}
  F\left(\mathbf{r},\rho\right)\equiv\frac{2}{\sqrt{\pi}}\sum_{\mathbf{l}}
  e^{-|\mathbf{l}+\mathbf{r}|^{2}\rho^{2}},
\end{equation}
where the vectors $\{{\bf l}\}$ correspond to the $N$ points of a
two-dimensional lattice of area $A$. $F$ is a periodic function of
$\mathbf{r}$, with the periodicity of the lattice. Therefore, it can
be expanded into the following Fourier series
\begin{equation}
  F\left(\mathbf{r},\rho\right)=\sum_{\mathbf{g}}F_{\mathbf{g}}
  \left(\rho\right)e^{i\mathbf{g}\cdot\mathbf{r}},
  \label{Fdef}
\end{equation}
where $\{\mathbf{g}\}$ are the reciprocal lattice vectors, and
\begin{eqnarray}
  F_{\mathbf{g}}\left(\rho\right) & = & \frac{2}{\sqrt{\pi}}\cdot\frac{1}{A}
  \int d^{2}r\,\sum_{\mathbf{l}}e^{-|\mathbf{l}+\mathbf{r}|^{2}\rho^{2}}
  e^{-i\mathbf{g}\cdot\mathbf{r}}\nonumber\\
  & = & \frac{2}{\sqrt{\pi}}\cdot\frac{N}{A}\int d^{2}r\,
  e^{-|\mathbf{r}|^{2}\rho^{2}-i\mathbf{g}\cdot\mathbf{r}}\nonumber\\
  & =&\frac{2\sqrt{\pi}}{A_{u.c.}\rho^{2}}
  e^{-|\mathbf{g}|^{2}/4\rho^{2}}.
  \label{appeniden}
\end{eqnarray}
Here, $A_{u.c.}=A/N$ is the area of a unit cell. Using
Eqs. (\ref{Fdef})-(\ref{appeniden}) and the identity
\begin{equation}
\frac{1}{\left|{\bf l}+{\bf r}\right|}=
\frac{2}{\sqrt{\pi}}\int_{0}^{\infty}d\rho\, e^{-\left|{\bf l}+{\bf r}\right|^{2}\rho^{2}}
\end{equation}
we obtain
\begin{eqnarray}
  \sum_{\mathbf{l}}\frac{1}{\left|\mathbf{l}+\mathbf{r}\right|}
  & = & \frac{2}{\sqrt{\pi}}\sum_{\mathbf{l}}\int_{0}^{\infty} d\rho \,
  e^{-|\mathbf{l} +\mathbf{r}|^{2}\rho^{2}}\nonumber\\
  & = & \frac{2\sqrt{\pi}}{A_{u.c.}}\sum_{\mathbf{g}}\int_{0}^{G} d\rho \,\frac{1}{\rho^{2}}
  e^{-|\mathbf{g}|^{2}/4\rho^{2}+i\mathbf{g\cdot r}} \nonumber \\
  & & +\frac{2}{\sqrt{\pi}}\sum_{\mathbf{l}}\int_{G}^{\infty} d\rho \,
  e^{-|\mathbf{l}+\mathbf{r}|^{2}\rho^{2}},
\end{eqnarray}
where the integral was split into two at an arbitrary positive value
$G$.  Finally, calculating the integrals leads to
Eq. (\ref{eq:ewaldquote}), where the divergent piece of the original
sum is given by the $\mathbf{g}=0$ term (and when ${\bf r}=0$ also the
${\bf l}=0$ term) in the new representation. The remaining part of the
infinite sums over $\mathbf{g}$ and ${\bf l}$ is rapidly converging at
a rate which is optimized by an appropriate choice of $G$.


\begin{thebibliography}{999}
\bibitem{3dslabs} It was argued\cite{Nagaev01}, however, that the combined
effects of a long-range Coulomb interaction, polarizability, and the
interaction of electrons with acoustical phonons may stabilize a
slab structure in three dimensions.
\end{thebibliography}
\end{document}